# GSTR: Secure Multi-hop Message Dissemination in Connected Vehicles using Social Trust Model

Anirudh Paranjothi[1], Mohammad S. Khan[2], Sherali Zeadally[3], Ajinkya Pawar[4], David Hicks[4]

[1]School of Computer science, The University of Oklahoma, – Norman, Oklahoma, USA.
[2]Department of Computing, East Tennessee State University, – Johnson City, Tennessee, USA.
[3]College of Communication and Information, University of Kentucky, – Lexington, Kentucky, USA.
[4]Department of Electrical Engineering and Computer Science, Texas A&M University, -Kingsville, Texas, USA.

*Abstract*— **The emergence of connected vehicles paradigm has made secure communication a key concern amongst the connected vehicles. Communication between the vehicles and Road Side Units (RSUs) is critical to disseminate message among the vehicles. We focus on secure message transmission in connected vehicles using multi-hop social networks environment to deliver the message with varying trustworthiness. We proposed a Geographic Social Trust Routing (GSTR) approach; messages are propagated using multiple hops and by considering the various available users in the vehicular network. GSTR is proposed in an application perspective with an assumption that the users are socially connected. The users are selected based on trustworthiness as defined by social connectivity. The route to send a message is calculated based on the highest trust level of each node by using the node's social network connections along the path in the network. GSTR determines the shortest route using the trusted nodes along the route for message dissemination. GSTR is made delay tolerant by introducing message storage in the cloud if a trustworthy node is unavailable to deliver the message. We compared the proposed approach with Geographic and Traffic Load based Routing (GTLR), Greedy Perimeter Stateless Routing (GPSR), Trust-based GPSR (T-GPSR). The performance results obtained show that GSTR ensures efficient resource utilization, lower packet losses at high vehicle densities.**

## INTRODUCTION

Vehicular Ad hoc NETworks (VANETs) [1] are used for communication between vehicles and also between vehicles and roadside infrastructures (e.g., Road-Side Units (RSUs)). Governmental transportation agencies and automobile manufacturing companies along with standardization bodies (ITS, ETSI, and so on) have been working together in the last few years to improve road safety and traffic management by leveraging VANET technologies. In the connected vehicle infrastructure, the vehicles create a network to share data directly with each other, a mode of communication often referred to as Vehicle-to-Vehicle (V2V) communication. Advances in the field of VANET are significant because they can address some of the increasing concerns such as the number of road accidents, safety, efficiency, traffic management and social connectivity between the sender and receiver.

The data exchanged in VANET is often sensitive, and therefore security is one of the critical requirements of the system [2,3]. As the driver's behavior depends on the integrity of the messages received, the network's security and topology will be affected drastically if any malicious user alters the messages in the network. Such attacks could cause traffic jams, false positioning attack, denial-of-service, and so on.

The various security concerns in VANET can be resolved by the proposed global security architecture [4]. This architecture consists of five levels of security, namely, material level, authentication level, trust level, message level, and cryptographic level. The trust level addresses all the security issues in the connected vehicle system. A trust model helps in identifying a trustworthy node from malicious nodes. The authentication level adds another level of security by authenticating nodes in the vehicular system.

The ease of use of social networking and the increased level of connectivity anywhere, anytime from any device have led to a rapid increase in its usage. The most commonly used social networks among people are 1) Facebook, 2) Twitter, 3) Instagram, and so on. Among all of them, Facebook users have been growing tremendously. As of 2019, Facebook has 2.32 billion users, an 11% increase year over year [5], which would help in finding the trustworthiness of a user for transmitting messages in the multi-hop VANET environment.

The use of social networking in VANET has been increasing because of various reasons:

1) Users with social network connections communicate with each other frequently while traveling. Such frequent communications can improve factors such as delivery ratio, latency, efficiency and so on of the routing protocol.

2) As social networking connections of users are initiated for long term communication and are less volatile, the social network-based connection's information does not frequently change and thus reduces the communication overhead.

3) By calculating a node's trust weight using a social network, the sender can find an optimized route to deliver the message to a connected node that is likely to move in the direction of the destination rather than blind forwarding [6,7]. Thus, we consider the use of social networking for message dissemination in VANETs to be essential.

The advent of Vehicular Cloud Computing (VCC) has enabled vehicles to establish a communication between the vehicles and to provide data transmission between the vehicles during the unavailability of connections in dense regions such as the Manhattan environment [8, 9]. In VCC, the cloud is used to store important statistical and messaging data. Moreover, the cloud ensures the safety and scalability of vehicles, which are essential factors when deploying smart and intelligent vehicular systems [10].

In this paper, we propose a trust model along with an authentication mechanism in conjunction with a social network (i.e., Facebook) for secure communications in a vehicular network. The vehicular network is assumed to be deployed in an urban area where the probability of meeting a connected vehicle using any social media is higher than that in the rural area. The trustworthiness of a message is calculated using the trustworthiness of the vehicle's user on a social network. The authentication of vehicle users is based on their social network connections. The more a node is socially connected and known to the other nodes, the higher is its trustworthiness. Once a node has connected to another node, it can connect to it faster in the future because of the implicit trust between them.

This rest of the paper is organized as follows. Section II presents related works on trust in VANET using social networks. Section III and IV describe the proposed architecture. Section V and VI present the proposed algorithm and performance evaluation results of our proposed approach respectively. Finally, we present the conclusion and future scope of the proposed approach in section VII.

## I. BACKGROUND AND RELATED WORK

With the growing interest in VANET, concerns related to the security of data in VANET have emerged, creating a new field of research. Various security solutions [11-13] for VANET have been proposed in the literature. The global security architecture proposed in [4] can be used as the basic building blocks for enhancing the security of VANET [4]. The proposed architecture consists of five levels: the hardware infrastructure, authentication, trust, data, and cryptographic operations.

The essential components of VANETs include 1) the Application Unit (AU), 2) the On-Board Unit (OBU), 3) the Road Side Unit (RSU) [14, 15]. The AU is a device on board which has applications provided by the service provider and communicates with the network of the OBU. The AU is usually connected to the OBU using a wireless or a wired link. The OBU consists of devices such as processors and resources on board to process data. The RSU helps in increasing the communication range between nodes by redistributing the data to the OBU. Fig. 1 depicts the basic VANET architecture and the types of communications in a VANET.

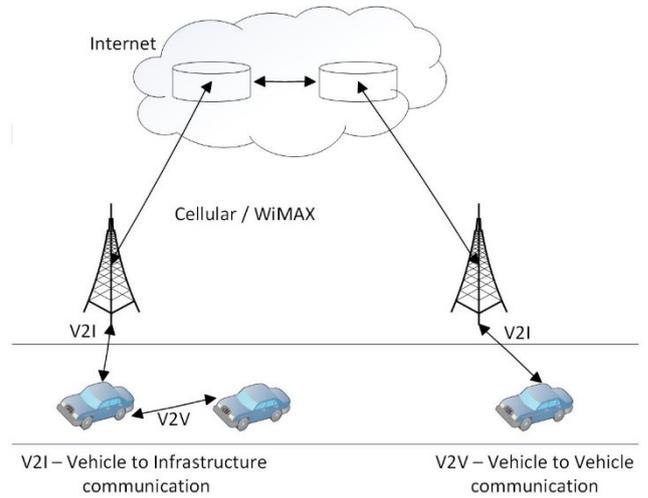

**Fig. 1.** VANET Architecture.

Communication in a VANET occurs in a single hop or through multiple hops. Fig. 2 depicts message passing in a multi-hop VANET system. In this paper, we propose a protocol which is based on multi-hop message transmissions. VANET shares many of the characteristics of a wireless network and the characteristics of an ad hoc network. The essential characteristics of VANETs are high mobility, dynamic nature, frequent disconnection, limited bandwidth and transmission power, availability of transmission medium, and so on.

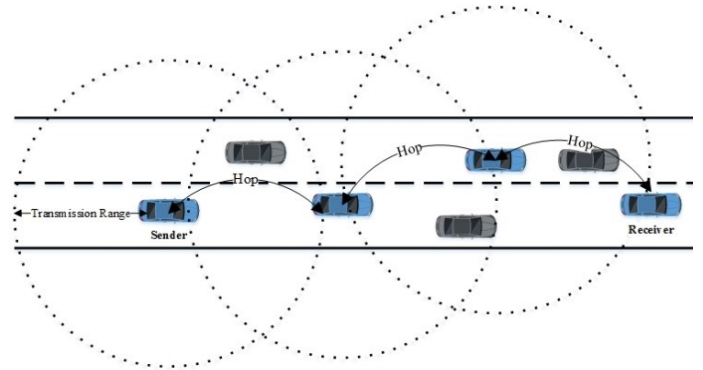

**Fig. 2.** Multi-hop message propagation.

Liao *et al.* [16] proposed a geographic multicast routing protocol for urban VANETs. The authors developed a Geographic routing based Social Dynamic Feature Aware (GeoSDFA) algorithm for routing the messages from a sender to the intended recipient. GeoSDFA not only considers the relevance of mobility and geography but also takes into account the vehicle's communication characteristics to destination areas. However, the limitation of GeoSDFA includes poor average delay and fewer delivery data when vehicles are unwilling to participate in data forwarding, or malicious nodes that have aggressive behaviors are present.

Chen *et al.* [17] demonstrate a high-level trust management model and its deployment scheme is based on a vehicular cloud system. The proposed model is a layered trust management mechanism that benefits from efficient use of physical resources (e.g., computing, storage, communication cost) and explores its deployment in a VSN scenario based on a three-layer cloud computing architecture. Moreover, performance modeling of the proposed trust management scheme is conducted through a novel formal compositional approach Performance Evaluation Process Algebra (PEPA). PEPA has superior features in compositionality and parsimony, which means that it can run efficiently model systems with layered architectures and complex behaviors. PEPA also supports various numerical analyses by calculating its underlying Continuous Time Markov Chains (CTMCs) directly or by solving a set of approximated Ordinary Differential Equations (ODEs). The authors also analyzed several key performance metrics such as end-to-end delay, delivery ratio, and others of the scheme and related capacity issues in its deployment.

Cui et al. have proposed a system model for authenticating users in vehicles by using certificates, which takes into consideration the public key but does not consider the user's current location and state [18]. Z. Huang et al. work on a similar principle (i.e., authenticating users in vehicles by using certificates in social networks). The proposed system requests opinions from the vehicles all around using voting to calculate the weight of a given node for a hop [10]. This was only a theoretical proposal and has not been implemented in a real environment.

Resnick et al. have described trustworthiness using a reputation system. The trust value is calculated in their system uses the following three rules: a) Information regarding trustworthy and untrustworthy peers; b) Encouraging peers to act trustworthy; c) Eliminating untrustworthy peers from the system [10]. However, the approach has long delays associated with it.

The importance of using cryptography to encrypt and decrypt VANET messages exchanged is considered by Mejri et al. [19]. The authors implemented confidentiality, integrity, and authentication in their proposal. In addition to that, they considered the aspects of security and routing. Routing includes information about various geographic routing protocols used in VANET to transmit the message from the sender to the receiver. Security includes the advantages and disadvantages of various encryption and decryption schemes used in VANETs.

In a multi-hop system [20], there are various techniques used for data transmission. The use of broadcasting the message to all of the nodes causes redundancy and data collision at the receiver. To mitigate this problem, the system proposed in [20] allows only a limited number of nodes to forward the message toward the receiver. This helps in reducing data flooding of the network [20,21].

Paranjothi et al. considered a model for authenticating messages using social networking [22]. The authors proposed a multi-hop system using a trust model to choose the node for message transmission. The trustworthiness of each node within the range of the sender is determined in order to find the next node for message dissemination. However, the approach will not work if the sender and receiver are not within the communication range of each other.

Whaiduzzaman et al. described the architecture of vehicular cloud computing and its services based on the network as a service, storage as a service, computation as a service and information as a service [23]. Rajput et al. presented an authentication protocol in VANET using cloud assisted conditional privacy. It uses a pre-issued identifier to authenticate vehicles and cloud-based certification authority. The proposed protocol makes use of storage as a service. The cloud is used to store the messages transmitted until a connection is established between the receiver and sender; however, as the cloud is used to store the large data delays associated with it.

Yao et al. [1] proposed an architecture that classifies trust level into three types: entity-centric (node-based) trust, data-centric trust, and combined trust. In the proposed architecture, the entity centric trust model calculates the trust level according to the weight assigned to each node. The authors classified the nodes into three types: 1) high level, 2) medium level, and 3) low level. High-level nodes are mainly referred to as the roadside infrastructures such as RSUs, base stations, and so on. The trust level of high-level nodes is always high because it ensures the Quality of Service (QoS) and high transmission range for vehicular communications. Medium level nodes are mainly referred to as public buses, ambulances, and so on. Low-level nodes are referred to as private cars, taxis, and others. The trust levels of medium and low-level nodes are always low because they are not suitable for long distance communications.

The difference between the GSTR and previous approaches [16-18] is the way trust calculated. GSTR uses social network connections of the nodes which help in the classification of connected nodes using their trustworthiness. Currently, the proposed system is the only system that can be used for multi-hop message dissemination using a social network based on trust calculation for the connected nodes.

**Contributions of this work**

We summarize the main research contributions of this work as follows:

1. We consider a vehicular social network with various base station scenarios for message dissemination among the vehicles using the social trust parameter.
2. In our proposed approach, GSTR is an application, which assumes that users are socially connected. This improves the scalability of GSTR.
3. We proposed a secure multi-hop message dissemination algorithm. The proposed algorithm was implemented in various experimental scenarios.
4. The experimental results obtained with our proposed algorithm demonstrate that it outperforms other previously proposed approaches for various performance metrics including message delivery ratio, numbers of hops and end-to-end delay.

## II. PROPOSED NODE SELECTION SCHEME

Until now the systems proposed in the literature [16-18] have relied on trust models for message passing. These models considered the pre-assigned weights for assigning a trust level to the nodes after considering the importance of the node as high for authoritative nodes and low for RSUs and so on [10,18,24]. The only shortcomings in these systems are the accuracy of the trust level of a node. A node may be ignored even though it is trustworthy due to its high delay and high packet loss ratio.

A VANET is a network which is highly dynamic and the connection time between two vehicles is too short of building trust amongst themselves [10]. Hence, the proposed system in this work addresses this drawback by leveraging social networking support to calculate the trustworthiness of a node.

QoS is an essential issue for inter-vehicular communications. To obtain a high QoS, the proposed system uses not only the trust model but also the authentication and cryptographic levels of the global security architecture [16] in VANET. The trustworthiness and authentication of the node can be achieved by using the social network connections of the user of the node.

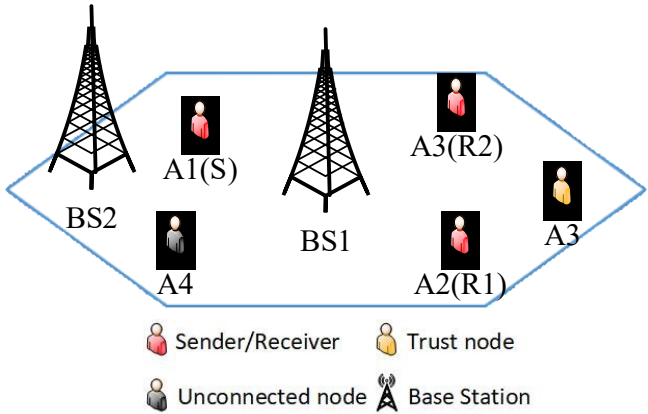

**Fig. 3.** Basic infrastructure for the proposed scheme

The proposed system uses multi-hop message passing in VANETs. In multi-hop message passing, a message sent from a sender node passes through multiple nodes before it reaches the receiver node. The main concern during these hops is data integrity. The data should be received intact as it was sent.

In our proposed system, the trust level identified the trusted nodes during the hops to transmit the data towards the receiver. Figure 3 shows the communication path from the source node to the destination node of the proposed system. The region under consideration is divided into cells with a base station at its center. All the vehicles are classified based on the weights assigned to each vehicle. The weights are calculated based on the trustworthiness of each node using social networks. The primary concern within such a system is to determine the direction of the vehicle under consideration. Message delivery to the receiver will fail if the node carrying the message is traveling in the opposite direction of the receiver node. This would cause message loss and system failure. To avoid such a failure, there should be an intelligent system that can prevent data loss and can handle the scalability of the system [10]. The message loss can be avoided by storing the message in the cloud when the node moves out of range of the base station. When a node connected to the receiver or the receiver itself comes into the range of the base station, the message is delivered.

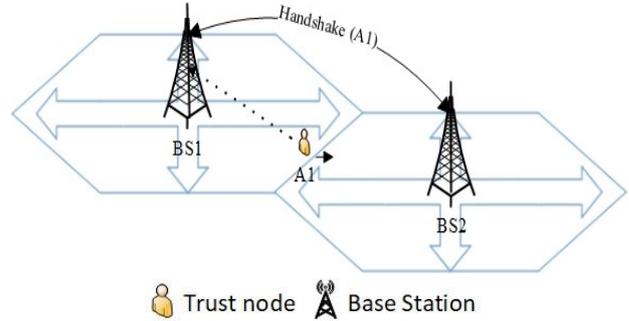

**Fig. 4.** The message returned from last base station to the home base station after a handshake between the two base stations along the multi-hop path.

As shown in figure 4, the trusted node carrying a message is about to leave the transmission range of base station 1 and enter into the range of base station 2. So, the base stations perform a handoff which switches from one base station to another. To avoid the message from the home base station from being lost and to prevent the message from getting further away from the receiver, the sending node sends the message to the base station which stores it in the cloud and that message is delivered to the receiver or a connected node when it comes in the range of the base station that received the message.

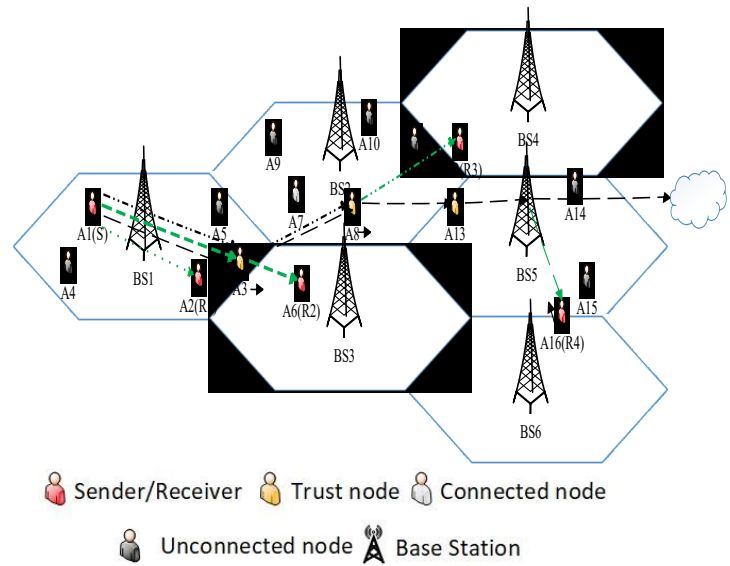

**Fig. 5.** GSTR multi-hop scheme.

Figure 5 illustrates the proposed scheme for multi-hop message delivery using a social network to build the trust model. The characteristics of each node used in the system are described below:

4) *Sender / Receiver Node*: The node which transmits the message is called the sender node. The node to which the message is to be delivered is known as the receiver node. When the receiver node is in the transmission range of the sender node, the system is said to be a single hop message delivery system. It does not need much authentication as there is no node in between the two.

5) *Connected Node*: The proposed system works on the principle of trustworthiness of a node. To calculate the trustworthiness of each node, we compute the weights using its connection between the sender and receiver nodes. In our system, we use Facebook as our social networking platform. If the node through which the message will pass through on its way to the receiver node is a "connected friend" of the sender and receiver, it is said to be a connected node. A connected node can also be termed as a node with the maximum number of connections between the sender node and the receiver node.

6) *Trusted Node*: A trusted node is a node which has the highest amount of trustworthiness. The trustworthiness of each node is calculated using the trust weight of each node. Trust weight, in our system, is the number of connections between the sender/receiver node and the connected node. Every receiver is a trusted node.

7) *Unconnected Node*: All the other nodes in the system, which are neither connected to the sender nor the receiver are called unconnected nodes.

Figure 5 shows the various nodes A1 to A16. A1 is the sender node and A2, A6, A12, and A16 are the various receiver nodes that are being considered. To make the system robust, message loss is prevented by leveraging a cloud infrastructure where messages are saved if a connected node is not available. To better understand the system, message delivery is explained using various cases:

Case 1: Message delivery when the receiver is in the sender's range.

Case 2: Message delivery based on the availability of a single connected node.

Case 3: Message delivery based on the availability of multiple connected nodes.

Case 4: Message delivery based on the unavailability of a connected node using storage in the cloud infrastructure.

As the messages are location oriented, we assume that each node carries a message only until it is in the range of the current base station. As it leaves the current base station's transmission area, it sends the message back to the current base station so that the receiver receives the message at the location it is intended for. The location of the sender node along with the social network connections creates various conditions. We consider each case individually below in order to understand the proposed system better.

**Case 1: Message delivery when the receiver is in the sender's range.**

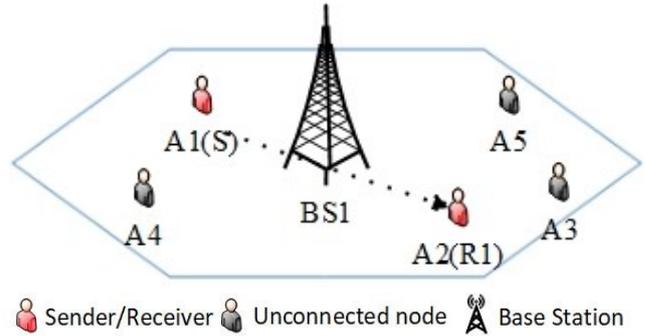

**Fig. 6.** Case 1: Message delivery when the receiver is in the sender's range.

As shown in figure 6, A1 is the sender and A2 is the receiver. Nodes A3, A4, and A5 are the unconnected nodes in the transmission range of A1 located in base station 1's cell. Unconnected nodes are the nodes which are not connected to the sender and receiver through the social network and hence are not considered trustworthy to deliver a message. When the sender A1 wants to deliver a message to the receiver, it searches for the receiver, A2, and discovers that it is within its transmission range. When it discovers that A2 is in the range of A1, A1(sender) directly sends a message to A2(receiver).

**Case 2: Message delivery based on the availability of a single connected node.**

In figure 8, A1 is the sender node, and A6 is the receiver node. Nodes A2, A4, and A5 are unconnected nodes in the range of base station 1 and are not considered. A3 is the only connected node between A1, the sender, and A6, and the receiver.

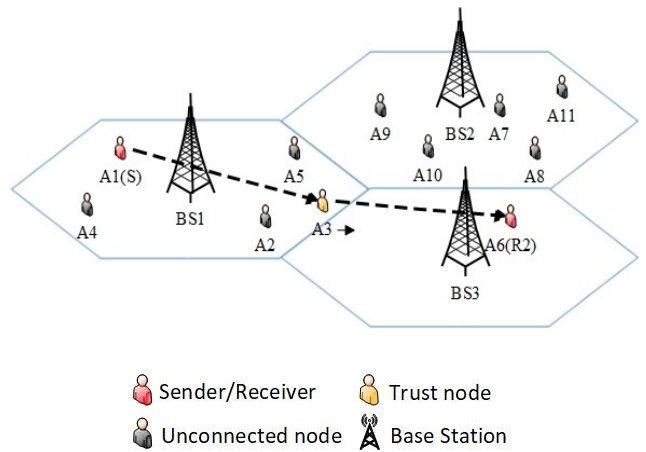

**Fig. 7.** Case 2: Message delivery from a sender to receiver in one hop through a trusted node.

Considering the availability of a single connected node, the connected node is assumed to be a trusted node, and the message is forwarded to A3 for delivery to A6. Furthermore, A3 searches for the receiver or a connected node between itself and the receiver. As shown in the figure, A3 has A6 in its transmission range and sends the message to A6 before it leaves the range of base stations 1.

**Case 3: Message delivery based on the availability of multiple connected nodes.**

Case 3 studies the trustworthiness of the connected node in a multi-hop environment. As shown in figure 8, A1 is the sender and A12 is the receiver in this case. A1 searches for a connected node between A1 and A12. A1 finds A3 to be the only connected node available and considers it as a trusted node to deliver the message to the receiver. In the next iteration of the algorithm, A3 becomes the new sender and searches for a connected node between A3 and A12.

**Fig. 8.** Case 3: Multi-hop message delivery in VANET using a social network for calculating trustworthiness.

Sender / Receiver  Connected Node  Trust Node  Unconnected Node

**Fig. 9.** Case 4: Message storage in the cloud infrastructure in multi-hop message delivery in a VANET using a social network.

Sender / Receiver  Connected Node  Trust Node  Unconnected Node

**Case 4: Message delivery based on the unavailability of a connected node using storage in the cloud infrastructure.**
As A3 discovers two connected nodes A7 and A8, it calculates the trust weight of each connected node to find a trusted node. The trust weight of each node is calculated using the connections they have in the social network. Let us assume that the trust weight of A7 to be 5 and A8 to be 9. As the connections between A3 and A8 are similar to the connections between A3 and A7, A8 is considered to be the trusted node and the message is delivered to A8 for further delivery to the receiver (A12). A8 becomes the new sender and searches for the receiver or a connected node. As A12 arrives in the transmission range of A8, A8 sends the message to the desired receiver (A12).

The system reduces message delivery failures by making use of cloud infrastructure in the system. The use of the cloud allows the user to save the message in the cloud if a connected node is not available. By storing the message in the cloud when a connected node is not available the delay increases, but it guarantees the message delivery without dropping the packets. In figure 9, A1 is the sender and A16 is the receiver. As stated in the above cases, the message delivery takes place through A3 and A8 when a connected node is available and by calculating the trust weight of the highest number of connections between sender and receiver. All other nodes, except for A13, are unconnected nodes. For this case, when A8 receives the message, it looks for a connected node and discovers A13 as a connected node and sends the message to A13.

A13 is the new sender node and searches for the receiver node. As the receiver node is not available, it searches for another connected node which may be close to the receiver. Unfortunately, there is no connected node between A13 and A16 in the base station 5 transmission range. When it does not find any connected node, it sends the message to the base station 5 (BS5) which stores it in the cloud infrastructure until it finds a connected node or the receiver which is in the range of the base station 5 (BS5). When the receiver A16, moving towards the direction of (BS5), arrive in the range of the BS5, the latter retrieves the message from the cloud and sends it to the receiver A16.

In all the above cases, the receiver node checks to ensure that a trusted node receives the message. If the message received is not from a trusted node, it eventually discards the message.

### III. Geographic Routing Protocol

The increase in the number of applications and the scalable nature of VANETs has made secure routing of data from a source node to a receiver node a challenging task. Most current research efforts on routing protocols for VANET [26-28] have been based on various existing routing protocols used in ad hoc networks. The most common used routing protocol in VANET is the Greedy Perimeter Stateless Routing (GPSR) protocol [1, 29]. GPSR uses greedy forwarding to send a message to the destination node along the shortest path available in the range and overcomes greedy forwarding failures by implementing perimeter search to find the node in the direction of the destination. However, GPSR has many performance drawbacks which include difficulties in performing better in urban environments in high vehicle density regions such as downtown regions and Manhattan environment, increase in overhead as the number of vehicles increases, and so on. [30]. Hence many researchers have proposed modified versions of GPSR a these include Greedy Perimeter Coordinator Routing (GPCR), Geographic Source Routing (GSR), Geographic and Traffic Load-based Routing (GTLR) [31]. All these modified versions of GPSR target urban areas.

Many of the routing protocols proposed for ad-hoc networks are inadequate and are not scalable for VANETs. This is because in VANETs, a dynamic the network topology is highly dynamic, and the number of available nodes also change frequently.

In this work, our proposed routing protocol is based on the principles of geographic routing protocols and uses a social network as a means to verify the trustworthiness of the nodes through which the message passes through on its way to the receiver node. Geographic Social Trust based Routing (GSTR) chooses a node that is socially connected to the sender and the receiver. The trust weight of each node is calculated using the social connections each node has. In the next section, we describe how we compute the weight of the trusted node and its connections.

A connected node with the highest trustworthiness is beneficial to the system only if the selected trusted node is traveling in the direction of the receiver from the sender. The system fails to deliver the message if the node is traveling in the opposite direction of the desired direction of the message to be disseminated. Hence, we need to consider the geographical aspect of the routing protocols. Current routing protocols search for the shortest route by choosing a node available between the given sender and receiver nodes. Similarly, the proposed GSTR includes the shortest path search along with the socially connected nodes to find the optimum path and avoid failures.

To understand the geographic routing better, we consider the following cases:

Case 1: The receiver is in the sender's range.

Case 2: A connected node is moving in the direction of the receiver.

Case 3: A connected node is not moving in the direction of the receiver.

**Case 1: Routing when the receiver is in the sender's range.**

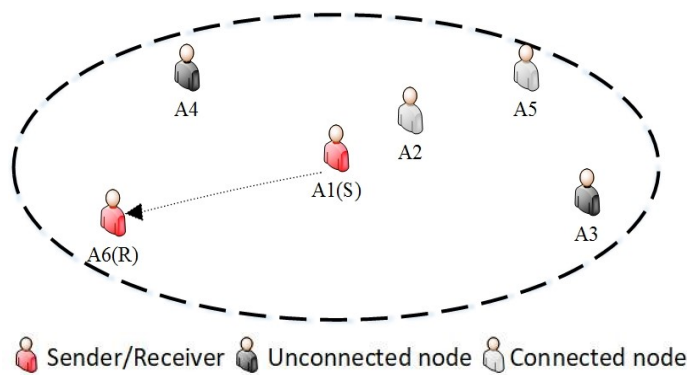

**Fig. 10.** Case 1: Routing when the receiver is in the sender's range.

When the receiver is in the sender's transmission range, the message transmission is a single hop from the sender to the receiver. The routing protocol does not consider the direction of the receiver while it stays within the sender's transmission range.

Figure 10 illustrates the first case in routing where the receiver is in the sender's transmission range. A1 is the sender, and A6 is the receiver. When the sender finds that the receiver in its range, the message is directly transmitted from A1 to A6 without taking into consideration the shortest path from the sender to the receiver as there is no multi-hop path between them.

**Case 2: A connected node is traveling in the direction of the receiver.**

In VANETs, vehicles communicate with each other through sensors and store data from the past to be used for the future. This data is essential for routing messages in the proposed system. In this case, the vehicle's direction of travel is used to determine if the connected node can be considered as a trusted node for message dissemination to the receiver. To calculate the movement of the node, the distance between the connected node and the final receiver node is constantly calculated in an aggregated manner. The distance decreases if the node is moving in the direction of the receiver, whereas the distance increases if the connected node travels away from the receiver node.

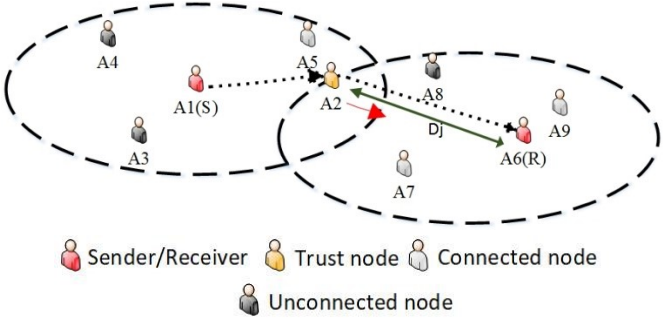

**Fig. 11.** Case 2: Routing when a connected node is moving toward the receiver.

As shown in Figure 11, the node A1 is the sender and wishes to send a message to the receiver node A6. Node A2 and A5 are the connected nodes in the range of the sender node. Based on the trustworthiness calculated (as explained earlier), the sender chooses the node A2 to be the trusted node. Another essential factor we need to consider is the geographic position and direction of the connected node (A2). As the arrow indicates the direction of A2 is towards the receiver, A2 is selected as the trusted node because it has the highest trust weight and is moving in the direction of the receiver. The direction of the node can be calculated by the past GPS path of the vehicle. As stated above, when the vehicle moves towards the receiver, the value of the distance $D_j$ decreases from its previous value.

**Case 3: A connected node is not moving in the direction of the receiver.**

Considering a similar scenario to case 2, when node A1 intends to send a message to node A6, A2 and A5 are the two connected nodes between the sender and the receiver. In this case, we assume that A2 has a higher trust weight than A5. Then, A2 is a stronger node to be trusted, but the direction of movement of the node A2 is opposite to the direction of the receiver. If the message sent to A2 must be delivered to A5, A2 being on another path will move out of range of the sender as well as the receiver. This will cause the message to be lost in the network thereby causing a message delivery failure. Hence, geographical tracking of the nodes becomes essential in this scenario.

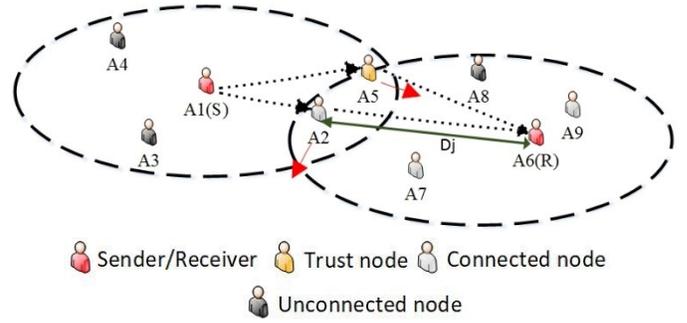

**Fig. 12.** Case 3: Routing when the connected node is not moving in the direction of the receiver.

As A2 moves in a different direction from the desired direction, the distance $D_j$ continues to increase. Taking this into consideration, sender node (A1) discards A2 as an option and considers another connected node (A5) to be a trusted node. Even though A5 has a lower trust weight than A2, but as it is moving in the direction of the receiver node, A5 is selected as a trusted node to send the message to the receiver (Figure 12).

Another failure may be caused if none of the connected nodes in the transmission range of the sender node move in the direction of the receiver. In this case, rather than allowing failure to occur, the message is sent to the base station and stored in the cloud until a connected node moves in the direction of the receiver or the receiver itself arrives in the base station's cell.

## IV. PROPOSED GSTR ALGORITHM

We studied the proposed system by using the two algorithms we have proposed namely, one for the sender side and one for the receiver side. The sender side of the algorithm consists of three main steps which include 1) extracting the topology, 2) searching for a trusted node and 3) storing the message in the cloud when the desired receiver or a connected node to the receiver is not available. The receiver side of the algorithm is divided into two processes namely, receiving a message from the sender or a connected node or receiving a stored message from the cloud via a base station. Both of the processes check for each other and the connected nodes in their respective friend list to manage the trustworthiness of each node.

**Trust model based on weights:**

Similar to most of the trust-based models [1, 2], the trust value is established for each node. The trust value in our system is evaluated using the connections in the social network for each node. The notion of weight is used in the system to depict the weight of each node according to its connections. As the VANET environment is highly dynamic, the following two conditions can be considered based on the previous interactions between the sender nodes. Case 1: if both nodes have interacted previously, i.e., the connected node is a previously trusted node; or case 2: if there was no previous interaction between the nodes, i.e., the connected nodes will communicate for the first time.

Now, the trust weight of each node helps decide on the message delivery for both the cases. We also consider a third case where no connected node is found. This helps to make the system robust because in this case, rather than discarding the message, the message is stored in the cloud infrastructure until the recipient or a connected node comes in the range of a base station to receive the message.

$$W_N x = \begin{cases} 1 & x = H \\ 0.5 & x = M \\ 0 & x = L \end{cases} \quad (1)$$

The value of the trusted node is determined by using the number of connections the node has in a social network that is common to the receiver and the sender. To add intelligence into the system, the weight of the nodes is first calculated using a prior connection. If a node does not have a previous connection, the node's trust value will be assigned as 0. If a node has previously connected to the sender and the receiver and is in the range of the sender, the node's weight is the highest. Hence, the 'high level' nodes with the trust weight of 1 are nodes that previously connected to the sender and the receiver. Having such a node in the range helps the message delivery to be faster and reduces the required search for connected nodes. As the system compares the connected node's trust value based on its connection, the next level (medium level) is a node previously connected to sender and receiver but, not in the range of each other. Hence, the 'medium level' node receives the trust weight of 0.5 to make our system robust; we include the cloud infrastructure to save the message until a trustworthy node is found. The value of $W_N x$ is 0 when there is no connected node in the sender's range.

**Node selection using trust weight:**

In a multi-hop message delivery environment, it is difficult to decide which node should be selected for the next hop. The proposed algorithm chooses the most trustworthy node for the next hop. Amongst the connected nodes, the node with the highest number of social network connections with the sender and receiver is considered a trustworthy node and is designated as a trusted node.

$$\alpha_i = \max_{i \to N}(Cn_i) \quad (2)$$

Equation (2) selects a connected node as an option for the trusted node. α selects the connected node which has the highest trust weight amongst all connected nodes. The trust weight of each node is calculated using its previous history and social network connections in equation (2). $Cn_i$ is the list of all the connected nodes in the sender's range.

**Node Selection based on the geographic shortest path:**

When selecting a path from the source to the destination, based on the availability of multiple connected nodes, the selection of a trusted node becomes crucial. Selecting a node with the highest trust value but moving away from the sender and the receiver will cause the message to be dropped in the network thereby resulting in a message delivery failure in the proposed system.

**Table 1.** Notations.

| Parameter | Description |
|---|---|
| $D_{n_i}$ | Closeness from the connected node to the receiver node at an instant $i$ |
| $D_i$ | Distance from the sender to the receiver node |
| $D_j$ | Distance from the connected node to the receiver node |
| $\beta_i$ | Node selection value based on the geographical location of the node |
| $Tn$ | The trusted node under consideration |
| $Cn$ | The connected node under consideration |

Hence, we need to select a node which is highest in trustworthiness and is geographically on the route to the receiver. To search the shortest route, we choose a connected node with the highest trust weight and at the same time moving towards the receiver.

$$D_{n_i} = \frac{D_i}{D_j} \quad (3)$$

$TN_{select}$ selects the node with high trust value to disseminate the message from the sender to the receiver. The value of $D_n$ at an instant $i$ determines if the connected node is to be selected as a trusted node. If the current value of $D_n$ is more than its previous value, then the connected node is not selected as a trusted node. However, if the current value of $D_n$ is lower than its previous value, then the connected node is selected as the trusted node for message dissemination to the receiver. As stated in equation (4) the sender selects the node in the direction of the receiver. The value of β is set to 1 if the node is moving in the direction of the receiver, i.e., the distance between the connected node under consideration is getting closer to the receiver, else it is set to 0 if the connected node is moving away from the receiver. Partly connected nodes are the nodes that have a trust value of 0.5, and connected nodes are the nodes that have a trust value of 1.

$$\beta_i = \begin{cases} 1; & D_{n_i} < D_{n_{i-1}} \\ 0; & D_{n_i} > D_{n_{i-1}} \end{cases} \quad (4)$$

All the values obtained in equation (2) and (4) will be used further to select the trusted node.

**GSTR: Geographic Social Trust Routing**

The selection of the trusted node is done by considering the trust weight and the direction of motion of the connected node. Equation (5) expresses the trusted node selection process. The value of $W_N x$ sorts the nodes in the transmission range into past trusted nodes, connected nodes, and unconnected nodes. This helps in searching for a connected node in a sequence from the highest trustworthy node to the unconnected node.

$$TN_{select} = \begin{cases} W_N x = 1 & AND & \begin{array}{l} \beta_i = 1, \quad Tn = Cn \\ \beta_i = 0, \quad Tn \neq Cn \end{array} \\ W_N x = 0.5, \text{Select}(\alpha_i) & AND & \begin{array}{l} \beta_i = 1, \quad Tn = Cn \\ \beta_i = 0, \quad Tn \neq Cn \end{array} \\ W_N x = 0 & & Tn \neq Cn \end{cases} \quad (5)$$

Upon selection of a connected node or a partly trusted node, sender checks $\beta_i$ for the direction of the node. From equation (4), if the node moves in the direction of the receiver, the connected node is considered as the trusted node. If the node is an unconnected node, no trusted node is selected, and the message is sent to the base station for cloud storage. Whenever a connected node is in the range of the base station, a trusted node is chosen using the above process, and the message is sent to the trusted node.

**Sender side algorithm:**

The sender side algorithm consists of three necessary steps: extracting the topology, searching for a trusted node and storing of the message in the cloud if no connected node is found.

Friend_List (i) is a list of all the friends connected between the sender and the receiver through the social network. Extract_Topology() is used to find specific details such as location, sender, receiver, and so on. Similarly, search_conn(i,j) searches for connected nodes between i (sender) and j (receiver). Highest(i) chooses the highest trust value node amongst the connected nodes searched.

| Algorithm_Sender (msg, loc, rec_name, conn_node, Friend_list(i), flag=0) |
|---|
| 1  **while** (flag != 1) |
| 2      Extract_Topology () |
| 3      **for all** j ∈ Friend_List (i) **do** |
| 4          recipient_exists = validation (j) ‖ |
| 5          conn_exist = validation(j) |
| 6          current_loc = location (loc) |
| 7          **if** (recipient_exists **AND** current_loc ==1) **then** |
| 8              Send (msg, rec_name) |
| 9              flag = 1 |
| 10             **goto** 31 |
| 11         **end if** |
| 12         **else** |
| 13             conn_exist=search_conn(send_name,rec_name) |
| 14             **if** (conn_exist **AND** current_loc==1) **then** |
| 15                 **if** (conn_node == prev_trust_node **AND** conn_node_dist(t) < conn_node_dist(t-1)) **then** |
| 16                     trust_node = prev_trust_node |
| 17                     **goto** 29 |
| 18                 **else if** (count (conn_node) == 1 **AND** conn_node_dist(t) < conn_node_dist(t-1)) **then** |
| 19                     trust_node = conn_node |
| 20                     **goto** 29 |
| 21                 **end if else** |
| 22                 **else** |
| 23                     trust_node = Highest(conn_node) |
| 24                     **goto** 29 |
| 25                 **end else** |
| 26             **end if** |
| 27         **end else** |
| 28         Send (msg, base_station) |
| 29     Send (msg, trust_node) |
| 30     sender_node = trust_node |
| 31  **end for** |

After extracting the topology and searching the friend list for connected nodes between the sender and the receiver, if no connected node is found, the data is sent to the base station to be stored in the cloud. Although the data is stored in the cloud, a connected node to the receiver or the receiver itself is continuously searching for the connected node in a recursive loop. When a connected node becomes available, the data is transmitted to it.

After a connected node is found (line 11), the sender determines if the connected node is a previously trusted node so that it can decide if the message can be sent through it. Otherwise, it searches for a connected node with the highest trust value. Upon selection of a trusted node, the message is sent to the trusted node for delivery to the receiver node (line 7). Figure 13 shows the flowchart for the algorithm at the sender side.

**Receiver side algorithm:**

In the receiver side algorithm, the message is delivered to the receiver in the following two ways:

1) Message from the sender or a connected node.
2) The message stored in the cloud from the base station.

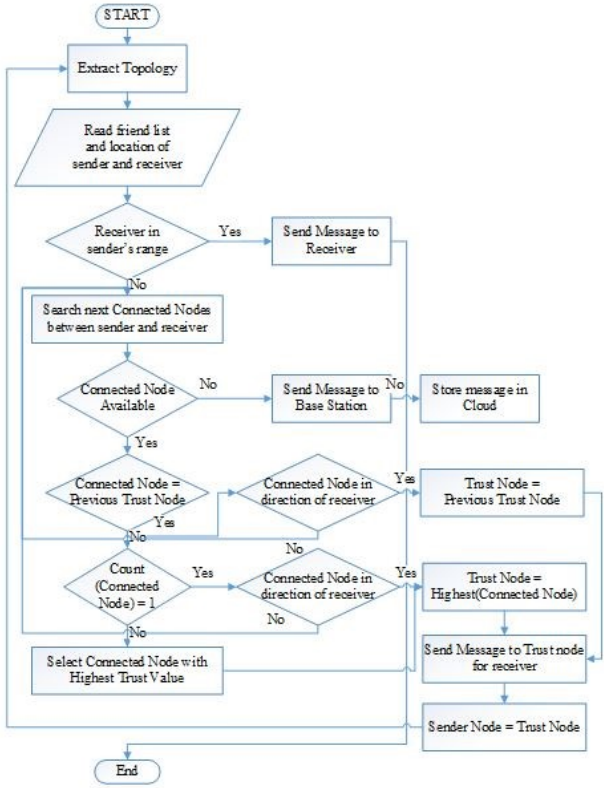

**Fig. 13.** Flowchart for the algorithm at the sender's side in the trust model using social networks for multi-hop message delivery in VANET.

To receive messages only from a trustworthy source, the receiver checks if the sender is present in the receiver's Friend_List(). To authenticate the sender, if the sender node is flagged as a trusted node, then we accept the message else we discard the message. Figure 14 describes the algorithm at the receiver side.

| **Algorithm**_Receiver (msg, loc, send_name, Friend_list(i)) |
|---|
| 1    **for all** j ϵ Friend_List (i) **do** |
| 2        sender_exists = validation (j) |
| 3    **end for** |
| 4    **if** (sender_exists **AND** sender == trust_node) **then** |
| 5        receive (msg, send_name) |
| 6    **end if** |
| 7    **else** |
| 8        discard(msg) |
| 9    **end else** |

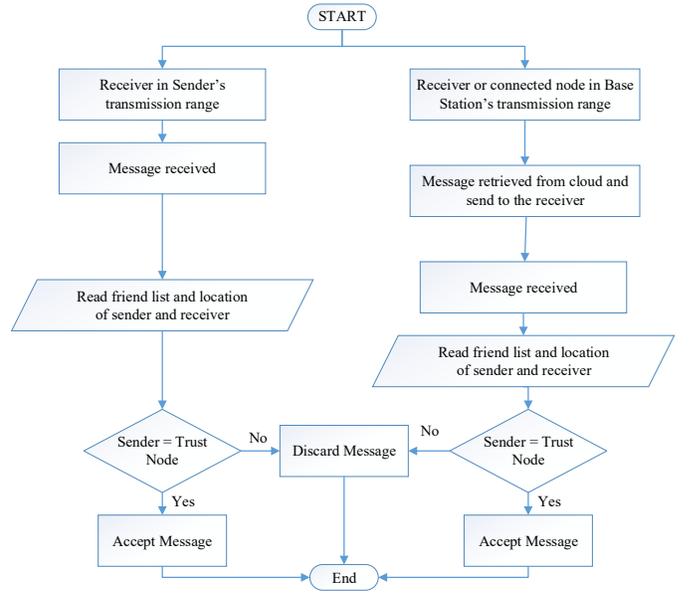

**Fig. 14.** Flowchart of the algorithm at the receiver's side in the trust model using social networks for multi-hop message delivery in VANET.

## V. SIMULATION RESULTS

The performance of our proposed architecture depends on the successful message dissemination from the sender node to the receiver node multi-hopping through an optimum route. Rather than using the normal voting or calculation of the node using the trust weights from the neighbors, the message delivery uses the social networking trust. In contrast to various existing systems, our proposed architecture works better on message dissemination because it delivers all types of messages instead of only emergency messages. The message dissemination is through trusted nodes which are determined by using social networking which also puts less load on the network as compared to broadcasting techniques used by various other systems [10].

We use the following performance metrics to evaluate the performance of GSTR and to compare our results with Geographic and Traffic Load based Routing (GTLR), Greedy Perimeter Stateless Routing (GPSR), and Trust based GPSR (T-GPSR) protocols.:

1. *Message Delivery Ratio*: Message delivery ratio is the ratio of the number of packets that successfully arrive at the destination to the total number of packets sent by the transmitter.
2. *Number of Hops*: The number of hops a message takes to travel from the sender to the receiver.
3. *End-to-End Delay*: The delay caused during the message dissemination from the sender to receiver.

*a)* **Analysis of the proposed system for various cases**

We studied our proposed system for various cases by considering the availability of a connected node in the sender's range. The following table summarizes the case study of our proposed system.

**Table 2.** Analysis of the proposed system for various cases.

| Factors \ Case | Message delivery ratio | Number of hops | End-to-end delay |
|---|---|---|---|
| Case 1 | Single connected node available | Single connected node available | Single connected node available |
| Case 2 | Multiple connected nodes available | Multiple connected nodes available | Multiple connected nodes available |
| Case 3 | No connected node available | No connected node available | No connected node available |

**End-to-end delay**

End-to-end delay is the delay caused during the dissemination from the sender to the receiver. We transfer the message using the nodes to reduce the combined delay to send the message from the sender to the cloud or RSU and then delivering it to the destination. Hence, case 3, where there is no connected node available and the message is stored in the cloud, incurs the highest average end-to-end delay.

However, message dissemination through the nodes on the same route of the sender and receiver is a faster mode of message delivery because the overhead incurred to deliver the message is reduced. Hence for case 1 and case 2 when the message is delivered either by a single connected node or multiple connected nodes, they have the least end to end delay.

**Message Delivery Ratio**

Message delivery ratio is the ratio of the number of packets that successfully arrive at the destination (receiver node) to the total number of packets sent by the transmitter (sender node).

The ratio of successful message delivery is higher in case 1 and case 2 because single or multiple connected nodes are available in the sender's range. The chances of successful message delivery increase even in the case of a node failure. However, when there is no node available in the sender's range (case 3), the message is stored in the cloud until the receiver or a connected node arrives in the desired destination area This does increase the delay in the message delivery but also assures a guaranteed message delivery thereby providing a higher QoS.

**Number of hops**

The average number of hops is the number of hops required for a message to be delivered from a sender to a receiver. Considering the number of hops, the higher the number of connected nodes available in the range, the higher the probability of finding a node closest to the receiver in the sender's range the smaller is the number of hops. The location of the connected node plays an essential role in determining the number of hops required for message dissemination.

In case 1, when a single connected node is available, the message has to be forwarded to the connected node irrespective of the geographical location of the node. The node might be too close to the sender, and consequently, there is an increase in the number of hops. However, in case 2, a node which is farthest from the sender, and within its range, towards the location of the receiver can be selected. This reduces the average number of hops.

In case 3, when no connected node is available, the message is stored in the cloud until a connected node or the receiver node is recognized in the desired location. This may or may not be the best solution as it can directly deliver the message to the receiver upon its arrival in the range, but the end to end delay of message delivery is the highest because the message is stored in the cloud rather than just forwarding it through other nodes.

*b) Simulation results comparing our proposed scheme with other existing routing protocols*

We evaluated the proposed system by comparing its performance with the performances of existing routing protocols which include: GTLR, GPSR, and T-GPSR. We consider the system's operation without any attacks in an urban area where the probability of meeting a connected node is high. The system can recognize previous trust nodes which will ensure faster communication between the connected nodes for message dissemination. We conducted simulation tests by varying the number of nodes from 40 to 200. The node's movement for the simulation was generated using Simulation of Urban MObility (SUMO) which provided the input to an NS-2 simulator.

**Case 1: Simulation results when a single connected node available.**

**Message Delivery Ratio:**

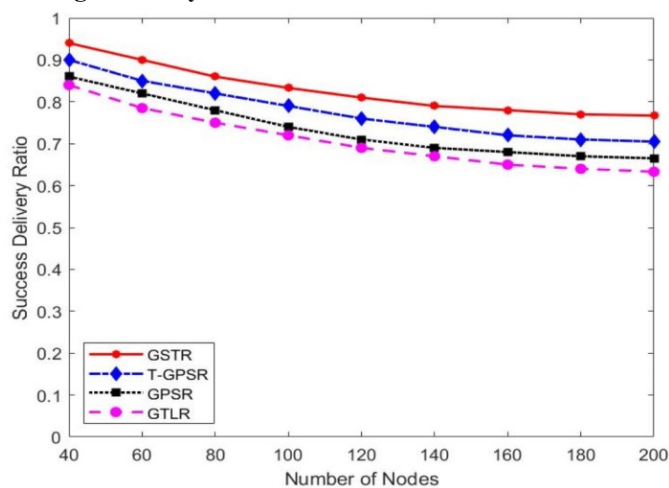

**Fig. 15.** Successful Message Delivery Ratio when a single connected node is available.

In GSTR, the message delivery ratio is based on the trustworthiness. However, when a single connected node is available the sender transmits the message directly to the connected node available (regardless of trustworthiness), which in turn transmits the message to the receiver when they are in the communication range of each other. Figure 15 shows the message delivery ratio when a single connected node is available. The message delivery ratio decreases as the number of users increases due to an increase in load on the network. However, GSTR performs better compared to GPSR, T-GPSR, and GTLR algorithms and provides more than 80% message delivery ratio for all vehicle densities when a single connected node is available.

**Average Number of Hops:**

In case 1, the sender transmits the message to the single connected node available for all vehicle densities. Thus, the average number of hops required to deliver the message is always 1.

**Average End-to-End delay:**

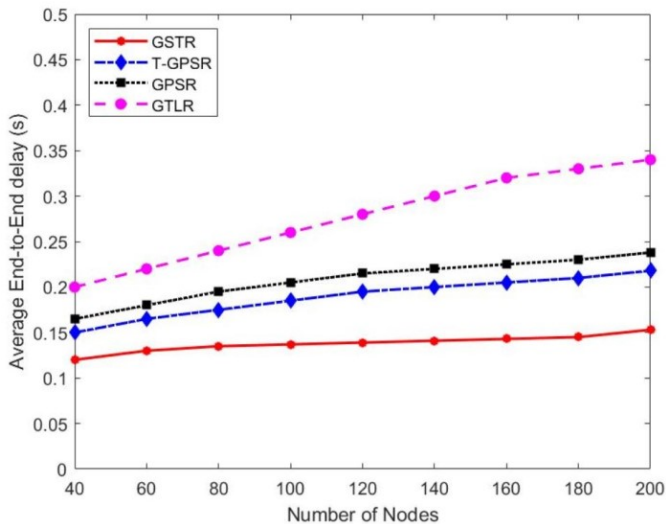

**Fig. 16.** Average End-to-End Delay when a single connected node is available.

When a single connected node is available, the average number of hops required to deliver the message is always 1. Thus, the average end-to-end delay is always less compared to when multiple connected nodes are available (case 2), and no connected node is available (case 3) at all vehicle densities. Figure 16 shows the average end-to-end delay when a single connected node is available. End-to-end delay of GSTR increases marginally when the number of users increases in the system because lots of messages need to be delivered in a specific time interval ($t$). GSTR outperforms GPSR, T-GPSR, and GTLR algorithms at all vehicle densities.

**Case 2: Simulation results when multiple connected nodes are available.**

**Message Delivery Ratio:**

The message dissemination from the sender to the receiver in the proposed system occurs by considering the trustworthiness of the connected nodes between the sender and receiver. The higher the trustworthiness of the intermediary nodes, the higher is the probability of successful message delivery to the receiver. Figure 17 shows the experimental results of the system's performance with the existing routing algorithms specified above when multiple connected nodes are available. GSTR performs better than any of the existing studied algorithms when we consider successful message delivery ratio performance metric.

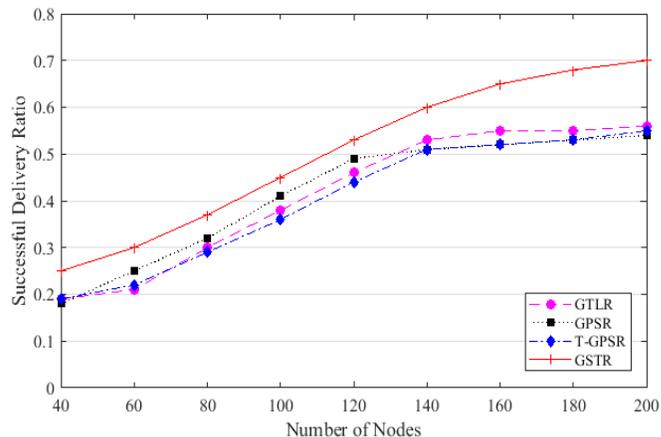

**Fig. 17.** Successful Message Delivery Ratio when multiple connected nodes are available.

As the number of nodes increases in the urban environment, the chances of spotting a socially connected trustworthy node increases. Hence, we note the increase in the successful message delivery ratio towards the end as the number of nodes increases.

**Average Number of Hops:**

An average number of hops evaluates some hops required to transfer the message when the sender and receiver are not within each other's transmission range.

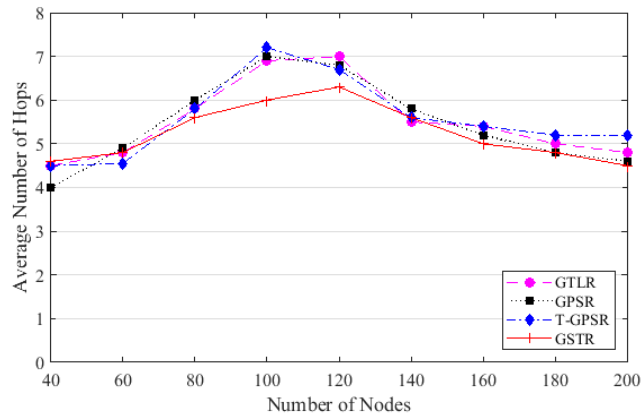

**Fig. 18.** Average Number of Hops when multiple connected nodes are available.

The number of hops affects the message dissemination delay. The fewer the number of hops, the faster the message dissemination. To use the least number of possible hops, the proposed algorithm considers the nodes in the location of the receiver node. Figure 18 shows a comparison between the existing and proposed algorithms when multiple connected nodes are available. The simulation results show a slightly higher average end-to-end delay when there are a few nodes available. However, when the number of nodes increases, the number of trust nodes in the direction of the receiver nodes increases. Hence, the average number of hops decreases when the number of available nodes increases.

**Average End-to-End delay:**

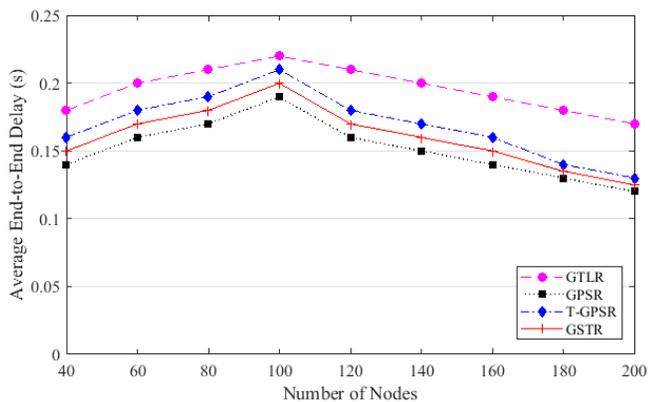

**Fig. 19.** Average End-to-End Delay when multiple connected nodes are available.

Based on the simulation results shown in figure 19, the proposed algorithm, GSTR, incurs a lower delay than the existing GTLR and T-GPSR algorithm. However, the end-to-end delay is marginally higher than that of GPSR. This is because the number of hops increases when fewer connected nodes are available on the system. However, when the number of users increases, the number of trust nodes increases in the system resulting in more messages delivered within a specific time interval ($t$).

**Case 3: Simulation results when no connected nodes are available.**

**Message Delivery Ratio:**

As we have discussed earlier, the message delivery ratio is based on the trustworthiness. However, when a trusted node is not available, the message is stored in the cloud infrastructure until a trusted node or the receiver node arrives which makes the system even more robust in terms of message delivery. This increases the end to end delay in message delivery but assures a guaranteed message delivery. Figure 20 shows the message delivery ratio when no connected nodes are available. GSTR yields a high message delivery ratio compared to GPSR, T-GPSR, and GTLR algorithms at all vehicle densities. This is because we use the cloud to disseminate message when no connected nodes are available.

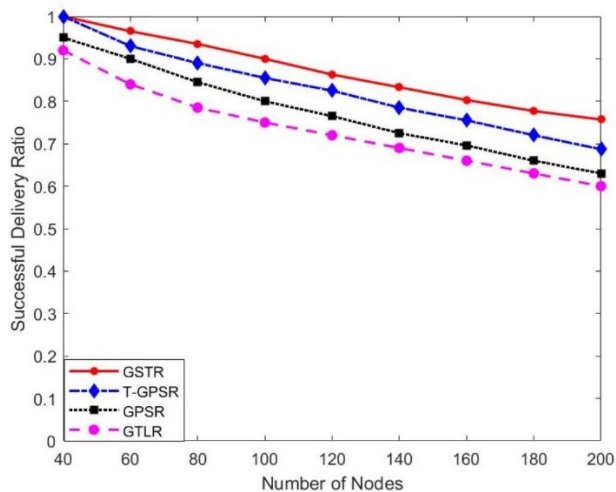

**Fig 20.** Successful Message Delivery Ratio when no connected nodes are available.

**Average Number of Hops:**

. When no connected nodes are available, the message is temporarily stored in the cloud to deliver it to the receiver. Thus, the average number of hops required to deliver the message is always 0 regardless of the vehicle densities.

**Average End-to-End delay:**

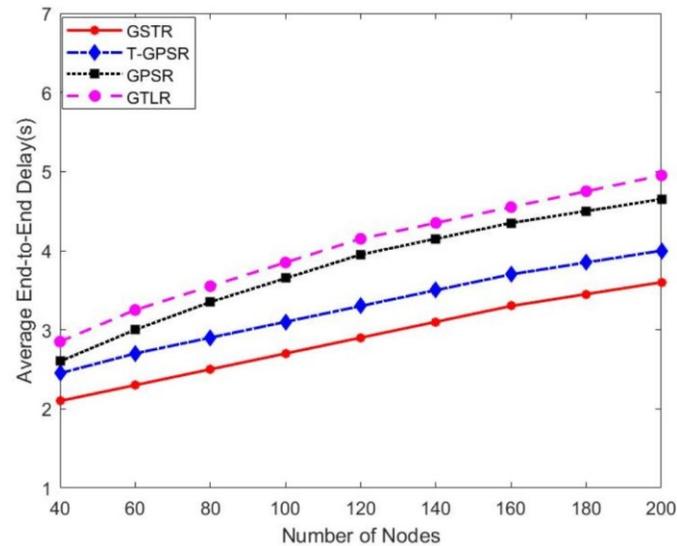

**Fig. 21.** Average End-to-End Delay when no connected nodes are available.

Figure 21 shows the average end-to-end delay when no connected nodes are available. As the connected nodes are not available to broadcast the message to the receiver, the message is stored in the cloud which results in the highest average end-to-end delay because of the additional communication overheads. However, GSTR has lower end-to-end delay compared to GPSR, T-GPSR, and GTLR algorithms at all vehicle densities.

## VI. Conclusion and Future Directions

In this paper, we have proposed a system for safe message dissemination using socially connected trust nodes. The usage of a cloud structure for storage of messages when a trusted node is not available helps to increase the chances of successful message dissemination. We use social network connections of the nodes to classify to determine their trustworthiness. To the best of our knowledge, our proposed system is the only system that can be used for multi-hop message dissemination using a social network based on trust calculation for the connected nodes.

We have analyzed the performance of our proposed GSTR and performed extensive simulations using the NS-2 and SUMO simulators. The results showed that GSTR is robust, efficient, and provides the best performance when compared to GTLR, GPSR, and T-GPSR protocols at high vehicle densities. Our proposed system results in fast and secure message dissemination using the minimum number of hops possible over the available trust nodes. It is worth pointing out that this system is targeted for the urban area where most of the nodes are socially connected. In this work, we have performed our experiments in a controlled simulation environment. As part of our future work, we plan to validate our simulation experiments in a real testbed environment. We also plan to investigate the changes that need to be made to the proposed system to support areas where there are few connected nodes.

### Acknowledgments
We thank the anonymous reviewers for their valuable comments which helped us to improve the content and presentation of this paper.

### References

[1] S. Zeadally, R. Hunt, Y. Chen, A. Irwin, and A. Hassan, "Vehicular Ad Hoc Networks (VANETs): Status, Results, and Challenges," Telecommunication Systems, vol. 50, no. 4, pp.217-241, 2012.

[2] X. Yao, X. Zhang, H. Ning, and P. Li, "Using a trust model to ensure reliable data acquisition in VANETs," Ad Hoc Networks, vol. 55, pp. 107–118, 2016.

[3] M. Raya and J.-P. Hubaux, "Securing vehicular ad hoc networks," Journal of Computer Security, vol. 15, no. 1, pp. 39–68, 2007.

[4] R. G. Engoulou, M. Bellaïche, S. Pierre, and A. Quintero, "VANET security surveys," Computer Communications, vol. 44, pp. 1–13, 2014.

[5] Zephoria valuable Facebook statistics website: https://zephoria.com/top-15-valuable-facebook-statistics/

[6] J. Shi, X. Wang, M. Huang, K. Li, and S. K. Das, "Social-based routing scheme for fixed-line VANET," Computer Networks, vol. 113, pp. 230–243, 2017.

[7] N. Vastardis and K. Yang, "Mobile Social Networks: Architectures, Social Properties, and Key Research Challenges," IEEE Communications Surveys and Tutorials, vol. 15, no. 3, pp. 1355-1371, 2013.

[8] T. Mekki, I. Jabri, A. Rachedi, and M. B. Jemaa, "Vehicular cloud networks: Challenges, architectures, and future directions," Vehicular Communications, vol. 9, pp. 268-280, 2016.

[9] S. Bitam, A. Mellouk, and S. Zeadally, "VANET-Cloud: A Generic Cloud Computing Model for Vehicular Ad-Hoc Networks," IEEE Wireless Communications Magazine, vol. 22, vo. 1, pp. 96-102, 2015.

[10] Z. Huang, S. Ruj, M.A. Cavenaghi, M. Stojmenocic, and A. Nayak, "A social network approach to trust management in VANETs," Peer-to-Peer network vol.7, no.3, pp 229-242, 2014.

[11] H. Hasrouny, C. Bassil, A.E. Samhat, and A. Laouiti, "Security risk analysis of a trust model for secure group leader-based communication in VANET," Vehicular Ad-Hoc Networks for Smart Cities, vol. 1, pp. 71-83, 2017.

[12] T. Oulhaci, M. Omar, F. Harzine, and I. Harfi, "Secure and distributed certification system architecture for safety message authentication in VANET," Telecommunication Systems, vol.64, no.4, pp. 679-694, 2017.

[13] J. Tellez, S. Zeadally, and J. Camara, "Security Attacks and Solutions for Vehicular Ad-Hoc Networks," IET Communications Journal, vol. 4, no. 7, pp. 894-903, 2010.

[14] A. Paranjothi, M. S. Khan, M. Nijim, and R. Challoo, "MAvanet: Message authentication in VANET using social networks," IEEE 7th Annual Ubiquitous Computing, Electronics & Mobile Communication Conference (UEMCON), pp. 1-8, 2016.

[15] Al-Sultan, M. M. Al-Doori, A. H. Al-Bayatti, and H. Zedan, "A comprehensive survey on Vehicular Ad Hoc network," Journal of network and computer applications, vol. 37, pp. 380-392, 2014.

[16] Q. Liao, Q, and J. Zhang, "Geographic Routing Based on Social Dynamic Features Aware in Vehicle Social Network," 15th International Symposium on Wireless Communication Systems (ISWCS), pp. 1-6, 2018.

[17] X.Chen, and L. Wang, "Exploring trusted data dissemination in a vehicular social network with a formal compositional approach," Computer Software and Applications Conference (COMPSAC), vol. 2, pp. 616-617, 2016.

[18] J. Cui, W. Xu, K. Sha, and H. Zhong, "An efficient identity-based privacy-preserving authentication scheme for vanets," International Conference on Collaborative Computing: Networking, Applications, and Work sharing, pp. 508-518, 2017.

[19] M. N. Mejri, J. Ben-Othman, and M. Hamdi, "Survey on VANET security challenges and possible cryptographic solutions," Vehicular Communications, vol. 1, pp. 53–66, 2014.

[20] W. Benrhaiem, A. S. Hafid, and P. K. Sahu, "Multi-hop reliability for broadcast-based VANET in city environments," IEEE International Conference on Communications (ICC), pp. 1-6, 2016.

[21] Y. Peksen and T. Acarman, "Multihop safety message broadcasting in VANET: A distributed medium access mechanism with a relaying metric," International Symposium on Wireless Communication Systems (ISWCS), pp. 346-350, 2012.

[22] M. Whaiduzzaman, M. Soohak, A. Gani, and R. Buyya, "A survey on Vehicular Cloud Computing," Journal of Network and Computer Applications, vol. 40, pp. 325-344, 2014.

[23] J. Zhang, "A Survey on Trust Management for VANETs," IEEE International Conference on Advanced Information Networking and Applications, pp. 105-112, 2011.

[24] H. Kaur, "Analysis of VANET geographic routing protocols on the real city map," Recent Trends in Electronics,

Information & Communication Technology (RTEICT), pp. 895-899, 2017.

[25] L. Liu, C. Chen, Z. Ren, T. Qiu, and K. Yang, "A Delay-Aware and Backbone-Based Geographic Routing for Urban VANETs," IEEE International Conference on Communications (ICC), pp. 1-6, 2018.

[26] C. Chen, Z. Wang, L. Liu, and J. Lv, "An Adaptive Geographic Routing Protocol Based on Quality of Transmission in Urban VANETs," IEEE International Conference on Smart Internet of Things (SmartIoT) pp. 52-57, 2018.

[27] A. Silva, K. N. Reza, and A. Oliveira, "An Adaptive GPSR Routing Protocol for VANETs," 15th International Symposium on Wireless Communication Systems (ISWCS), pp. 1-6, 2018.

[28] H. Li, A. Guo, and G. Li, "Geographic and traffic load based routing strategy for VANET in urban traffic environment," IET 3rd International Conference on Wireless, Mobile and Multimedia Networks (ICWMMN), pp. 6-9, 2010.

[29] B. Karp and H. T. Kung, "GPSR: Greedy perimeter stateless routing for wireless networks," 6th annual international conference on Mobile computing and networking, pp. 243-254, 2000.

[30] H. Li, A. Guo, and Guangyu Li, "Geographic and traffic load based routing strategy for VANET in urban traffic environment," IET Conference Proceedings, pp. 6-9, 2010.